# A Comparative Study of Milestones for Featuring GUI Prototyping Tools

**Thiago Rocha Silva[1], Jean-Luc Hak[1,2], Marco Winckler[1], Olivier Nicolas[2]**

[1]ICS-IRIT, Université Paul Sabatier, Toulouse, France
{rocha,jean-luc.hak,winckler}@irit.fr

[2]SOFTEAM/CADEXAN, Toulouse, France
{jlhak,onicolas}@e-citiz.com

## Abstract

Prototyping is one of the core activities of User-Centered Design (UCD) processes and an integral component of Human-Computer Interaction (HCI) research. For many years, prototyping was synonym of paper-based mockups and only more recently we can say that dedicated tools for supporting prototyping activities really reach the market. In this paper, we propose to analyze the evolution of prototyping tools for supporting the development process of interactive systems. For that, this paper presents a review of the literature. We analyze the tools proposed by academic community as a proof of concepts and/or support to research activities. Moreover, we also analyze prototyping tools that are available in the market. We report our observation in terms of features that appear over time and constitute milestones for understating the evolution of concerns related to the development and use of prototyping tools. This survey covers publications published since 1988 in some of the main HCI conferences and 118 commercial tools available on the web. The results enable a brief comparison of characteristics present in both academic and commercial tools, how they have evolved, and what are the gaps that can provide insights for future research and development.

## 1. Introduction

Every project is unique in terms of the business and technical problems that arise, the priorities assigned to it, the resources required, the environment in which it operates, the culture of the stakeholders, and the project manager's attitude used to guide and control project activities. Nonetheless, a closer look on actual developer's activities will reveal that many iterations are often necessary to mature design ideas, to explore design alternatives, and to convince customers (both client and end-users); such observation contradicts a linear view of software development based on straightforward sequence of steps (such as waterfall approach). Development processes imply planning of activities that ultimately will transform client/customers requirement into products that fulfill user's expectations. Therefore, it is sensible to ask how to take into account users' needs along the development process.

The ISO 13407 standard for human-centered design processes for interactive systems [1] (also known as user-centered design – UCD process) tackle this issue by placing users at the center of the development process. Prototyping Graphical User Interfaces (GUIs) is considered as one of the most important activities in a User-Centered Design (UCD) process as a mean to investigate alternative design solutions. In early phases of the development process, paper and pencil mockups are a suitable alternative for prototyping user interfaces [2]. They are inexpensive and yet suitable presentation models that can be used to communicate basic ideas with users about the design. Although paper-based prototypes remains useful, the passage from paper to software is error-prone as paper-based prototypes are informal descriptions that can be subject to interpretations (i.e. ambiguity in the recognition of the graphical elements) and they might provide insufficient information to describe some design constraints (ex. precise size and position of objects).

It is worthy of recalling that prototypes are an important way to communicate and discuss requirements as well as usability and ergonomic aspects, in particular in early phases of the development process [3]. Low-fidelity techniques of prototyping help designers to sketch and to present new ideas and concepts about the user interface. In early phases of the development process, prototypes are useful to involve users in participatory design activities, where users can directly influence what is being designed. As the development process progresses, medium and high fidelity prototypes are useful to refine features. Advanced prototypes provide more accurate information of the design alternatives thus helping to make decisions. The need for support activities such as planning, sketching, designing and evaluating prototypes of user interfaces, has led to the development of specialized tools.

Since 1988, with Mirage [4], the most important conferences in the field of Human-Computer Interaction (HCI) have given space for tools developed in order to solve several scientific challenges related with this theme. However, dedicated tools for supporting prototyping activities only started to have an impact in the market by 2003. Thus, we can observe a temporal gap between the research interest and market adoption [5].

The aim of this work is to investigate the state of the art in GUI prototyping tools. We present a review of academic and commercial tools. Our main contribution lies on analyzing both academic and commercial tools in terms of new ideas and features, regarding the main milestones they introduced over time. With such analyses, we have identified their coverage levels for these milestones and provided a detailed classification for them looking for new research gaps in this area. The next section of this paper presents the research protocol used to investigate tools in both academic and commercial contexts.

## 2. Methodological Approach

The present study encompasses both academic and commercial tools. Most of academic tools we analyze have been developed as a proof of concepts to support claims raised by scientific research. Despite the fact that some academic tools might be considered very advanced prototypes, they rarely make a breakthrough towards the market. Conversely, we consider commercial tools those that have been developed for making money either by selling rights of use or by allowing others to make money using them (for free) to accomplish work in an industrial scale. We consider the analysis of commercial tools important because they are decisive to understand the adoption of features originally available in academic tools.

The analysis of tools followed four main steps: selection of tools, classification of tools, revision and identification of target milestones, and finally discussion of the findings. It was analyzed prototyping tools for drawing (intended to support generic interface drawings), sketching (intended to get a basic concept – sketch – of how the user interface will look like), and wireframing (intended to refine the concept of how the user interface will work, normally using pre-defined interaction elements) based on their capability to provide useful prototypes. The analysis of academic tools was mainly based on the review of the literature. For commercial tools, we have only analyzed those that are readily available for download on the web. Hereafter, we present a comparative analysis in Table 1 and further details in the next subsections about how academic and commercial tools were selected for the study and how they have been classified. The following keywords were used in the search of both academic and commercial tools: prototype, prototyping tool, prototyping interface, wireframe, wireframing, sketch, sketching, draws and drawing.

**Table 1.** Contrastive analysis of research methods for academic and commercial tools.

| Criteria | Academic tools | Commercial tools |
|---|---|---|
| **Selection source** | HCI conferences | Web |
| **Search keywords** | prototype, prototyping tool, prototyping interface, wireframe, wireframing, sketch, sketching, draws and drawing | |
| **Number of initial entries** | 8.682 | 118 |
| **Exclusion factors** | Domain-oriented conferences, tools not published as full papers, tools for specific environments, model-based prototyping tools for multimodal user interfaces, and tools available in other languages than English. | No free version available, tools that are not standalone software, tools no longer updated and documented, and domain-specific tools. |
| **Number of tools retained** | 17 | 104 |

### 2.1. Selection of Academic Tools

We sought top ranked HCI conferences and selected those that were sponsored or co-sponsored by ACM, IEEE and/or IFIP. We discard domain-oriented conferences (such as mobile, embedded, robot, pervasive and ubiquitous interfaces) and conferences whose proceedings are available in other language than English. The final list of conferences includes: ACM CHI (1982-2016), ACM UIST (1988-2016), ACM DIS (1995-2016), ACM EICS (2009-2016), and IFIP INTERACT (1984-2015).

At first, we have selected papers that contain in the title and/or abstract any of the keywords presented in Table 1. With these keywords, we have found 8.682 publications. Subsequently, we have screened the papers and excluded those reporting tools developed for specific prototyping in specific environments (e.g. sketches of buildings for architects, drawings for designers, circuits and physical devices for engineers and so on). We did not take into account papers reporting model-based prototyping of multimodal user interfaces because our main interest lies in tools that can support the concrete development of user interfaces, not only to model it. Finally, we only considered publications of full papers. We have also included ActiveStory Enhanced [21] in the list because, despite the fact that it was not published in the target conferences but in the XP International Conference, it describes features that we consider relevant for the discussion.

In total we have retained 17 tools as follows: SILK [6], DENIM [7], DEMAIS [8] and CogTool [9] (from ACM CHI), Gambit [10] (from ACM DIS), GRIP-it [11] (from ACM EICS), Mirage [4], Ensemble [12], Lapidary [13], Druid [14] and Monet [15] (from ACM UIST), SIRIUS [16], MoDE [17], SCENARIOO [18], Freeform [19] and SketchiXML [20] (from IFIP INTERACT), and ActiveStory Enhanced [21].

## 2.2. Selection of Commercial Tools

We have used Google search engine to find commercial tools that match the keywords shown in Table 1. Using the links provided by Google, we visited the corresponding web sites to check the availability of tools for download. Only tools that were free or have a free period of evaluation were retained for further analysis. We have also included in the analysis some tools such as PowerPoint and Photoshop. Despite the fact that these tools cannot be properly called prototyping tools, they are often reported as suitable alternatives for building low-fidelity prototypes.

In total, we retained 118 tools for a second round of inspection. We have analyzed the tools' main features and paid particular attention to the way they handle the creation of the user interface and the precision that can be achieved when describing the behavior of the prototype. The subsequent analysis sought to find answers for the following questions: "Is the tool a standalone software or an extension/library/framework?", "Is prototyping generic interfaces possible?", "Is there a free trial of the tool?", "Is the tool still updated and documented?", and "Does the prototype produced with the tool support any interaction?". We have also inspected the mechanisms available for specifying the presentation (i.e. the graphical elements) and the dialogue (i.e. the behavior) of the prototype.

We decided to exclude tools that only provide libraries and/or stencil themes for help with the drawing of paper-based prototyping. The same decision was applied to tools that were no longer updated/documented. Domain specific tools, such as for automotive, were also excluded. In total, we have retained 104 commercial tools as shown in Table 2. The analysis of these tools allowed us to classify them in three categories depending on what can be prototyped: the behavior, the presentation (visual aspect) or both. The first one gathers 10 tools that are more suited for representing the behavior of a prototype. In the second one, we have regrouped 9 drawing tools like Inkscape or Photoshop, where it is possible to create a visual prototype without caring about the behavior or the possible interactions. Finally, the last category corresponds to tools that can manage both graphical and behavior aspects and it features the remaining 85 tools. Therefore, we have decided to focus on this last category since they are mainly tools that are dedicated to the construction of fully functional prototypes.

### 2.3. Classification of Tools

Both academic and commercial tools were inspected and classified according to the following group of criteria: description of the tools, features for edition, execution, management and evaluation. For "description of the tools", we have catalogued information about version, offers available, dependencies, backup policies (including cloud), platforms for editing and running, integration with other tools, export of code and file formats, and finally generation of documentation and code. For "features for edition", we have investigated features related to the presentation and dialog edition such as notations, degrees of fidelity supported, how to build the internal and the external dialog, how to handle conditions, parameters and actions, support for annotations, reuse and management of design options, interaction techniques supported and visualization. Execution features have been evaluated for the dialog execution, including notations available to describe navigation between windows, simulation engine, etc., possibility of annotating the prototype in runtime and/or if alternative design views are provided. In addition, we have also evaluated if the tools supported embeddable wireframes. On the management side, we have looked for features to control and customize favorite libraries, as well as mechanisms to control versioning of prototypes. Lastly, we have investigated features for supporting evaluation during the design process such as means of collecting feedback from users and/or other designers and specific features for running usability and user testing.

### 2.4. Identification of Milestones

Lastly, we have inspected the tools enlisted in Table 2 looking for common characteristics and/or functionalities that tools implement over time. Such analysis brought the identification of milestones concerning interaction techniques used to build prototypes (ex. pen-based interaction, widgets) requiring or not programming skills. This analysis also revealed aspects of the user interface that could be prototyped (the presentation and/or the behavior), the support for collaborative work, code generation, usability testing and design through the whole lifecycle, as well as reuse mechanisms (ex. libraries, templates, modules, patterns), aspects of scenario management, including version control and annotations, and mechanisms for running prototypes. We have identified 13 milestones that we consider worthy of further discussion, including: Non-Programming Skills, Pen-Based Interaction, Widgets, Behavior Specification, Collaborative Work, Reuse Mechanism, Scenario Management, Preview Mode, Support for Usability Testing, Support for Code Generation, Version Control, Annotations, and Support for the Entire Design Lifecycle. These milestones are presented in detail in Section 3.

## 3. Presentation of the Milestones

This section presents the milestones in detail and it illustrates tools that present the features mentioned as milestones. Section 4 presents a broader discussion about the evolution of the tools and the coverage of milestones.

### 3.1. Non-Programming Skills

Non-Programming Skills refers to the possibility of building prototypes without any prior programming skills. The first prototyping tools appeared with the advent of User Interface Management Systems (UIMS) [22], which aimed at separating the process (or business logic) from Graphical User Interface (GUI) code in a computer program [23]. UIMS were aimed allowing designers and developers to build software without any programming skills. The ultimate goal of UIMS tools was to allow users to concentrate on what is to be done rather than how to do it [4]. One way to accomplish this objective was to give users the ability to directly manipulate the representations of concepts from the task domain (e.g. design objects). Examples of tools pursuing this goal include MIRAGE, Lapidary, Ensemble, DENIM and Druid.

Non-programming skills is a driving feature that motivate the research on End-User Programming tools [30] which are aimed at empowering users to create what they need (or at least define more precisely part of what they need). Non-programming skills is considered a milestone because most of tools that came after the first appearance of this feature does not require (much) programming abilities from users. For instance, DENIM [7] is a pioneer example that illustrates how tools can be used for involving users into the design of the web sites to be developed. Some exceptions exist such as Lapidary, for instance, demanding some Lisp programming ability to express more refined behaviors.

Nowadays, it is a common sense between developers of Graphical User Interface (GUI) tools that they should simplify the activity of designers and interface engineers, and requiring some level of programming skills is a throwback. Because of that, among all the tools analyzed, only those that are more focused on the modeling, instead of GUI prototyping, still require some kind of programming. All the others work with abstract elements and behavior models to provide prototyping resources for users, without requiring an ability to program the software. This is a well-established feature today.

### 3.2. Pen-Based Interaction

Pen-based interaction allows hand-written drawings, which some authors claim that this is an intuitive passage from paper-based prototypes to interactive (software-based) prototypes [6]. Prototyping tools that implement pen-based interaction allows designers to keep the habit of drawing the user interface by replacing paper and pencil by digital sketching. In order to remove the ambiguities of drawing, some prototyping tools such as SILK (see Figure 1) also implement sketch recognition, which allows interpreting drawings and transforming them into graphical elements (widgets in a higher level of fidelity) that can be reused for building incremental prototypes. At Figure 1, we can see at the right side the results of sketch recognition applied to a hand-written drawing (shown at the left side) using SILK.

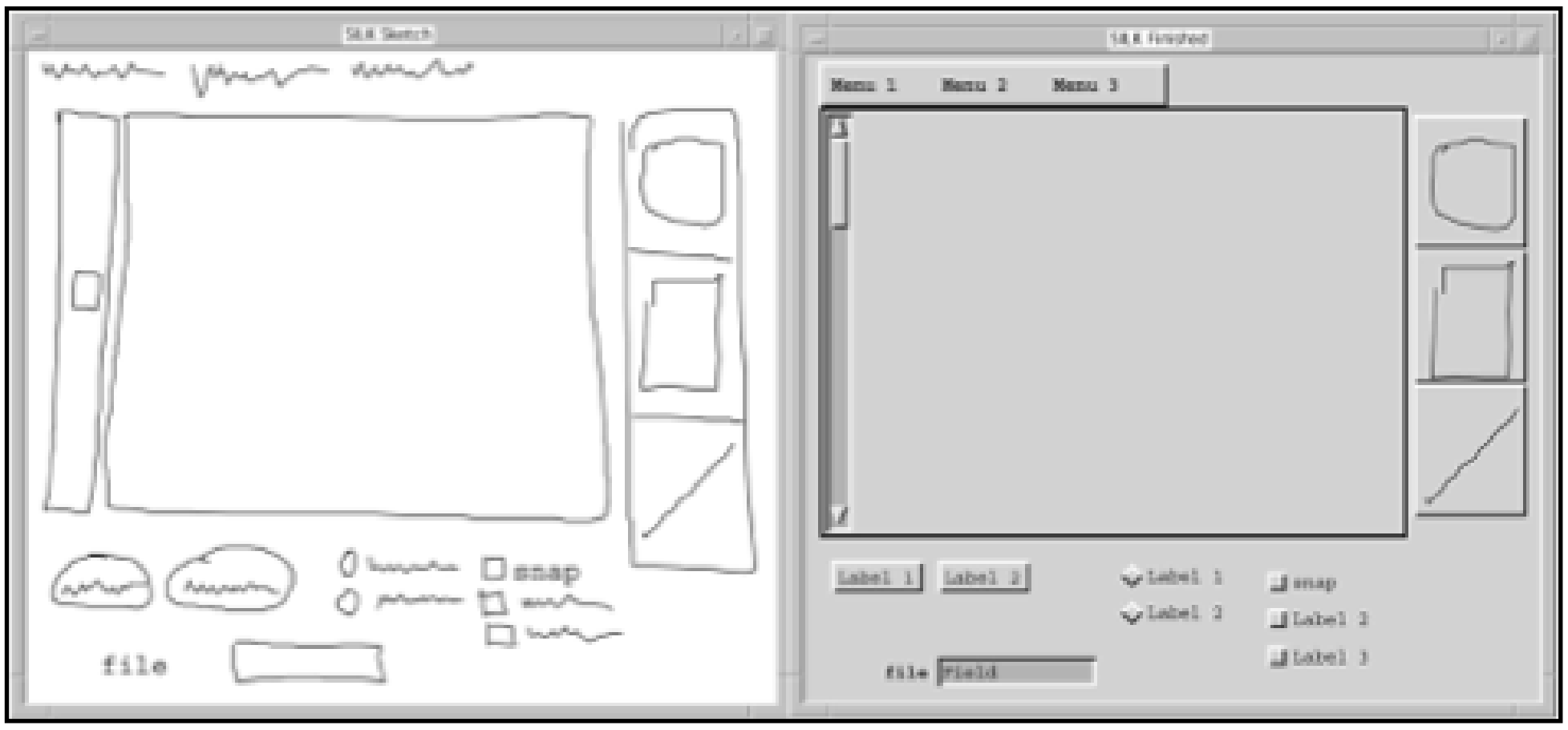

**Figure 1.** SILK. Left side: sketching widgets. Right side: transformed interface.

It is interesting to notice that few commercial tools implement pen-based interaction. Tools like Blueprint, Cacoo, Mockup Plus, NinjaMock and Pidoco, for instance, allow both pallet and sketching methods of interaction, but not sketching recognition.

## 3.3. Widgets

Widgets are pre-defined GUI elements (such as buttons, text fields, etc.) that users can interact with to perform their tasks with the user interface. Libraries of widgets are commonly available in prototyping tools and they were already in use in Lapidary. Their use guides the major part of tools that works with a palette as interaction technique nowadays. Widgets have the advantage of making the selection of graphical elements easier, offering a fast manner to set various components as menu bars, buttons, input form fields, and containers such as windows. It is interesting to notice that even tools that work with a sketching mechanism like SILK and DENIM have a library of widgets for common elements (drawn before) and treat them as a widget for future uses.

All dedicated prototyping tools we have analyzed present a library of widgets. The inner inconvenient of these libraries is that the palette is limited to a predefined set of components featuring widgets. Indeed, prototyping tools provide different level of look and feel and some provide full support of a clean layout of the components. For instance, Balsamiq (Figure 2) provide only rough design by claiming that they only focus on low-fidelity prototyping, whereas SceneBuilder for JavaFX provide components with a polished aesthetic and layout since it is designed for finished software. Other tools like SILK do not support directly the use of widgets to build prototypes, but they allow transforming the sketch made using pen-based interaction to real widgets, through sketch recognition techniques.

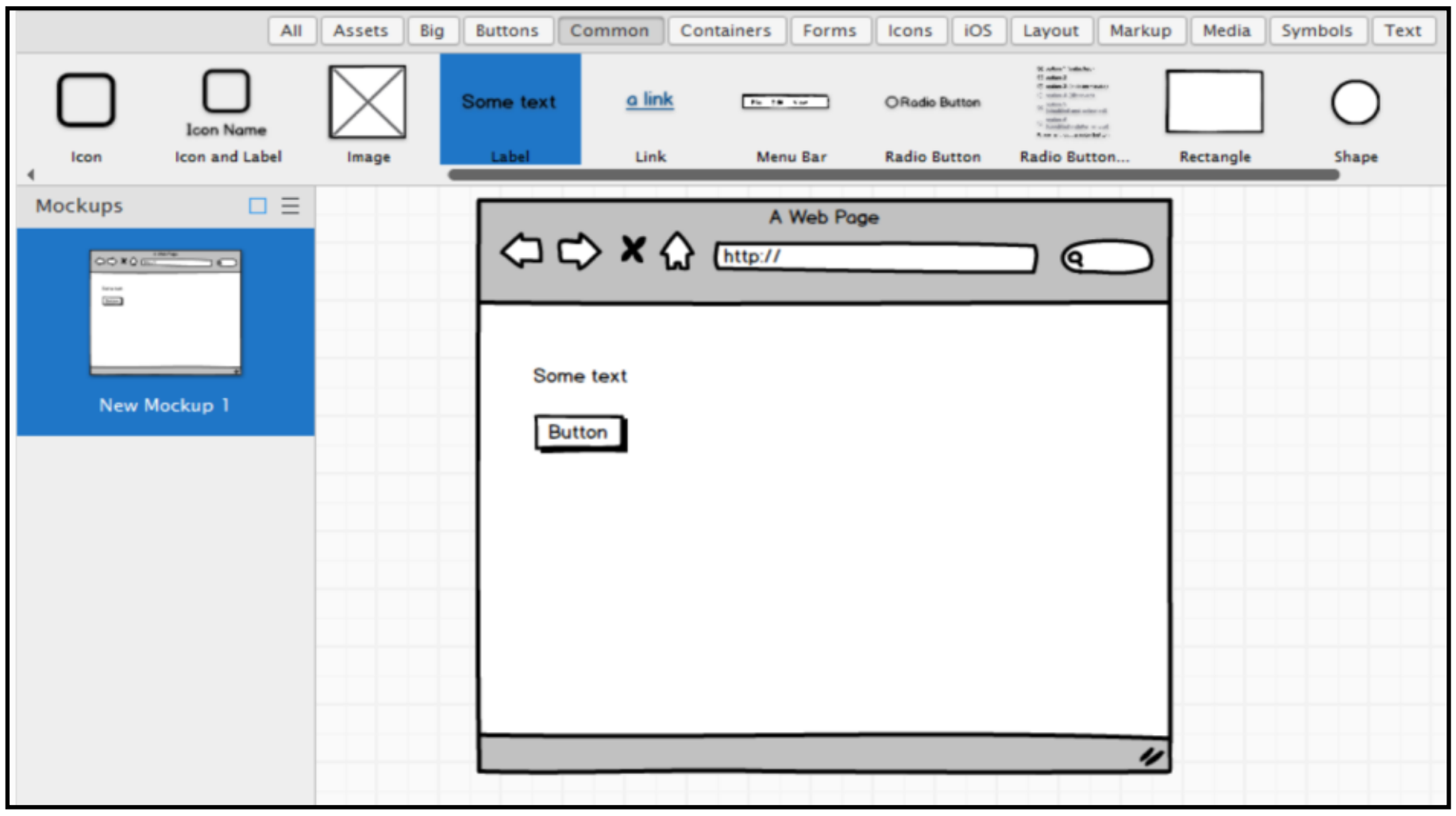

**Figure 2.** Example of tool that uses a palette of widgets (Balsamiq).

Many tools focus on the presentation by promoting the import of images to create high-fidelity prototypes like Origami or Atomic.io (Figure 3). Although those prototyping tools provide features to create basic shapes (i.e. rectangle, circle) and edit their properties (i.e. color, opacity, background image), those tools mainly emphasize their compatibility with drawing tools like Sketch or Photoshop. Prototyping tools provide also many options to animate widgets or create transitions between the different screens of the prototype. Tools like Invision or Atomic.io integrate a timeline to define the duration of the animation for each object in the prototype.

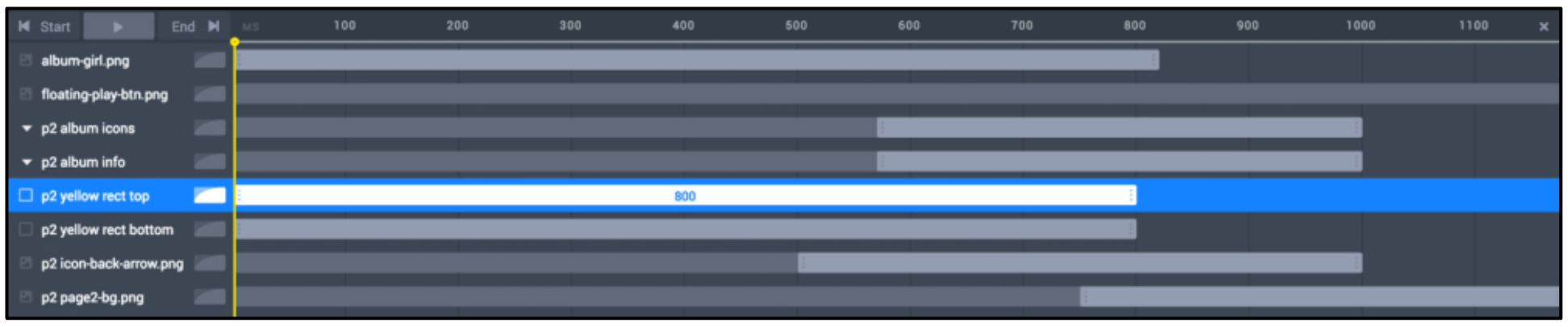

**Figure 3.** Animation timeline from Atomic.io.

Thus, those prototyping tools are used as a tool to organize assets between several pages to link those pages together and to share the prototype with others.

## 3.4. Behavior Specification

Behavior Specification refers to the ability to add dynamic behaviors to prototypes. Behavior is often described as a set of states that prototypes can reach by the means of transitions between states. Not all prototyping tools deal with behavior specification, many of them only allow to create static images of the presentation. As we shall see, there are many ways for specifying the behavior including setting hotspots on images, events handling on widgets and/or scripting in models.

Tools that employ hotspots allows the creation of areas on top of images (see Figure 4) that capture events triggered by the user. Designers need to create one hotspot for each part of the interface they want to make interactive. States are defined as static images of the prototype whilst transitions are associated to the hotspots on top of it. The problem with this method is that hotspots are associated to graphical areas without a particular semantics with the graphical element that is represented by the image.

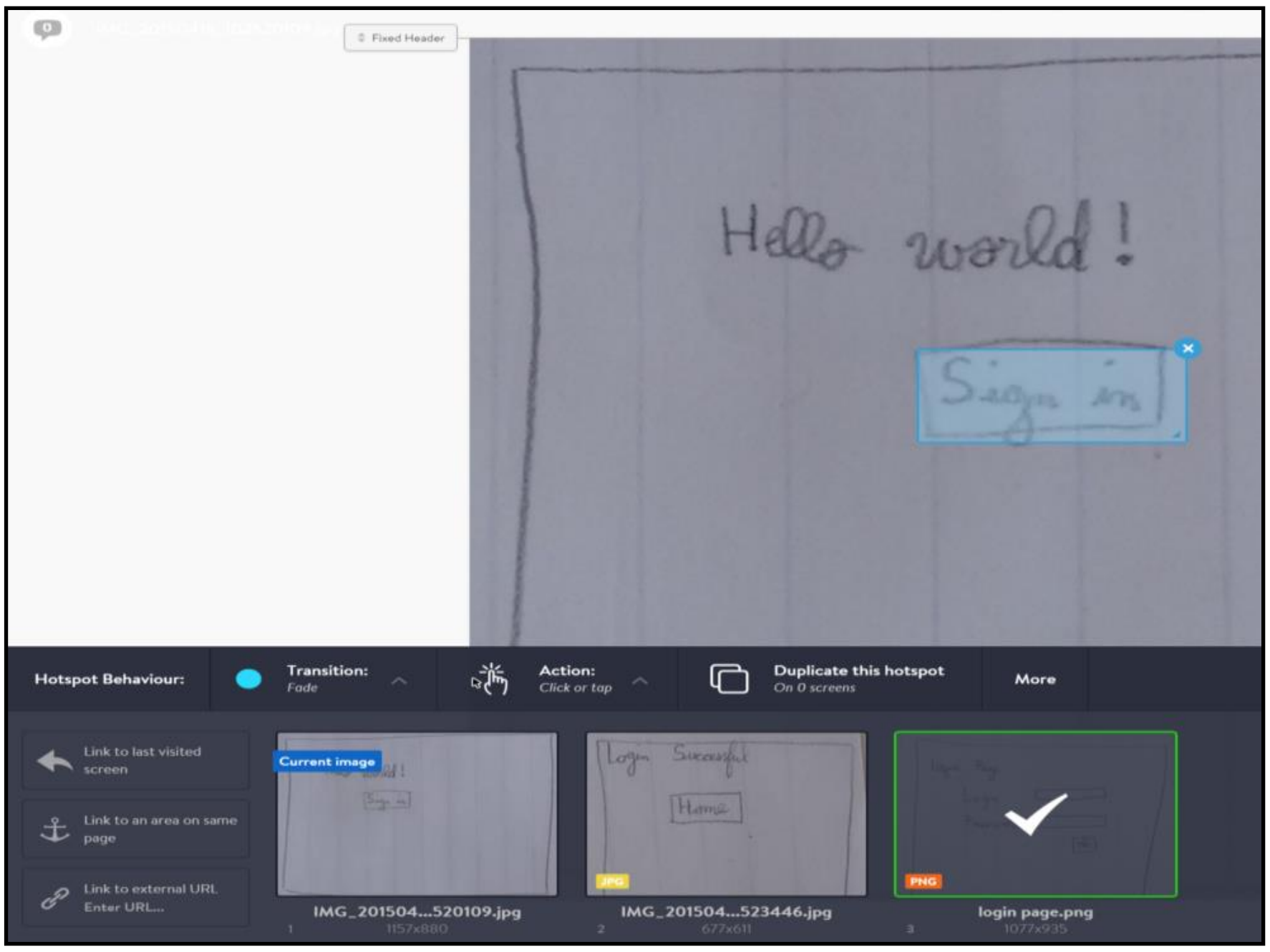

**Figure 4.** Example of hotspot using Marvel.

Unlike the tools using images for the presentation, wireframe tools uses widgets to build the interface. Therefore, wireframe tools generally do not require the use of hotspots since it is possible to create event handlers directly on the widgets (Figure 5). Each widget has a property "Event" that can be customized with the action required to trigger the event and the action that has to be made. By doing so, the dialogue is more dependent on the presentation.

In tools that describe prototypes as models, state machines and prototypes can be used to specify fine-grained behaviors. The behavior specification using models is often called dialog. One of the advantages of formally modeling the dialog is that it provides a computational mean to simulate the prototype behavior. Figure 6 shows the dialog for a "Login" application using the tool ScreenArchitect employing a state machine model for specifying the behavior. Notice the window "Login" on display in the foreground and the state machine specification in the background. That state machine indicates what will happens after the user has entered the login and password. This co-execution between the state machine and the presentation aspect of the prototype works in both ways: the state machine controls what is on display to the user who can trigger events that change the current state in the state machine. The prototype can be modified independently from the state machine to make it match with new requirements or feedbacks. One problem that arises when modifying the prototype and the state chart is that they are no longer consistent. Co-evolution is more expensive (in terms of workload) than just having a prototype to modify, but this allows having a formal description of the prototype behavior.

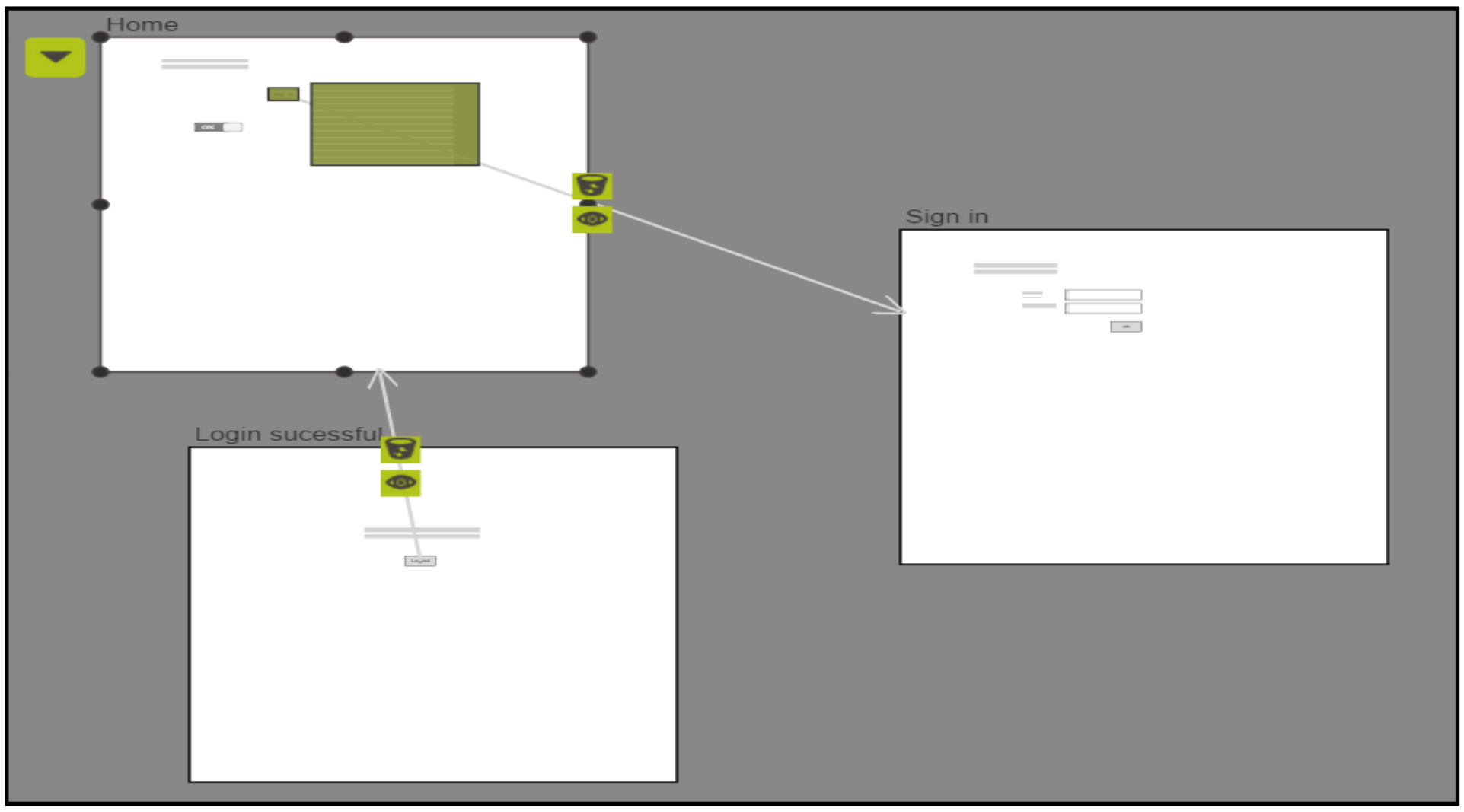

**Figure 5.** Events being handled in Pidoco.

As far as methods for specifying behaviors are at a concern, almost all of academic tools provide some kind of behavior specification. Lapidary was the first one we noticed. Interesting resources were provided after it, leading to the dialog construction. Unlike other dedicated tools, ActiveStory Enhanced, Balsamic, SILK, DENIM and Pencil, for instance, support only basic wireframe interactions, with links between screens and state changing. Tools like AppSketcher, Axure, CogTool and JustInMind are already able to specify conditions, editing properties or using variables, while Appery.io, JBart and ScreenArchitect support programming code as well.

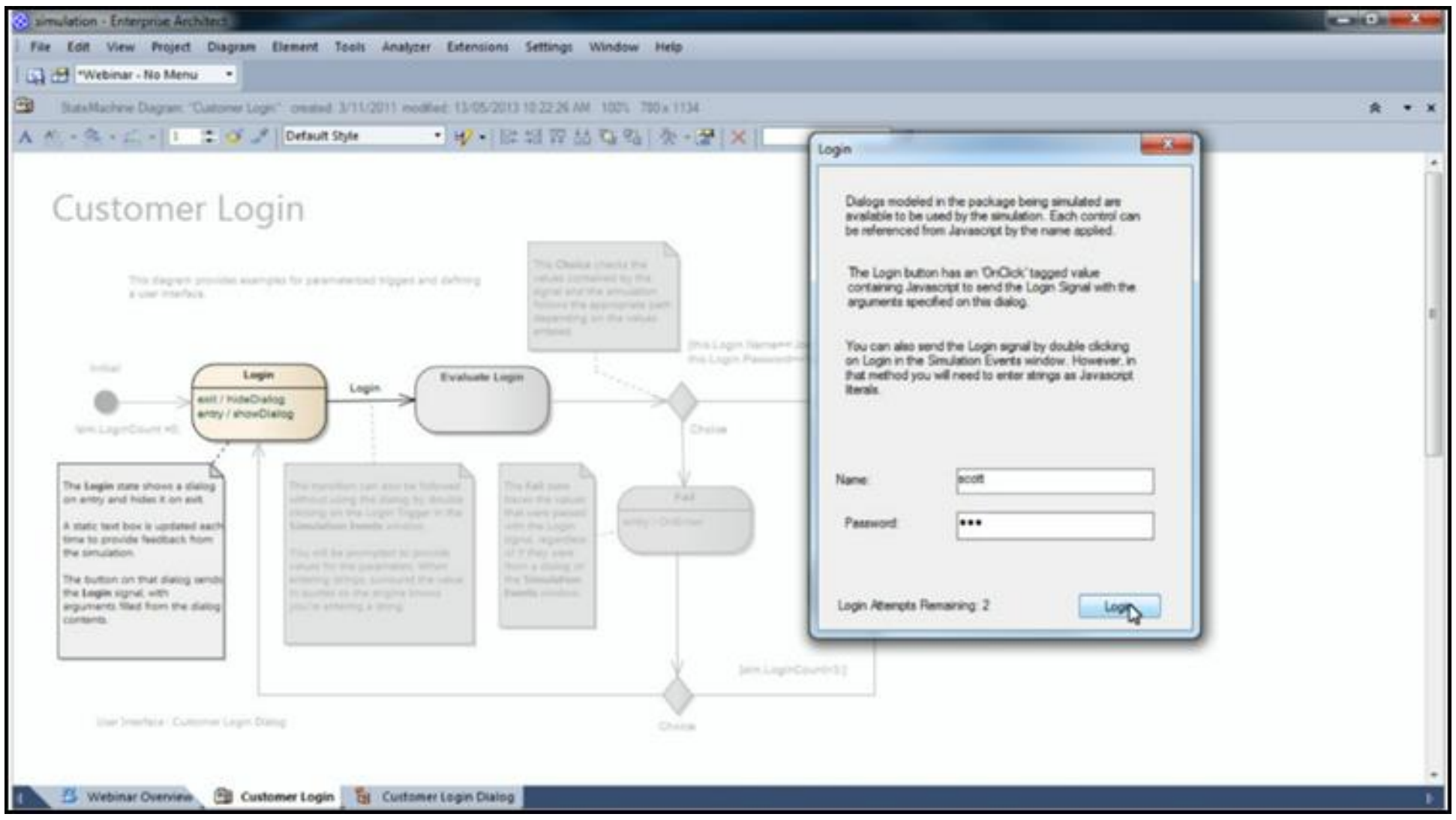

**Figure 6.** The state machine and the prototype associated with the state "Login" in ScreenArchitect.

## 3.5. Collaborative Work

Collaborative work refers to the support that allows people to work together (synchronously or asynchronously) on the same prototype. This is one to the most recent features in prototyping tools. Sangiorgi et al. [10] highlight that existing software for UI design by sketching shares the same shortcomings: only one person at a time can sketch a UI on one device or computing platform at a time with little or no capability for sharing sketches. Gambit is one of the few prototyping tool that supports collaborative work. Gambit implements many collaborative features such as collaborative creation and visualization of sketches on different devices, management of private and/or public mode with broad views of the drawings (like papers arranged on a wall) and a fine view of them (Figure 7).

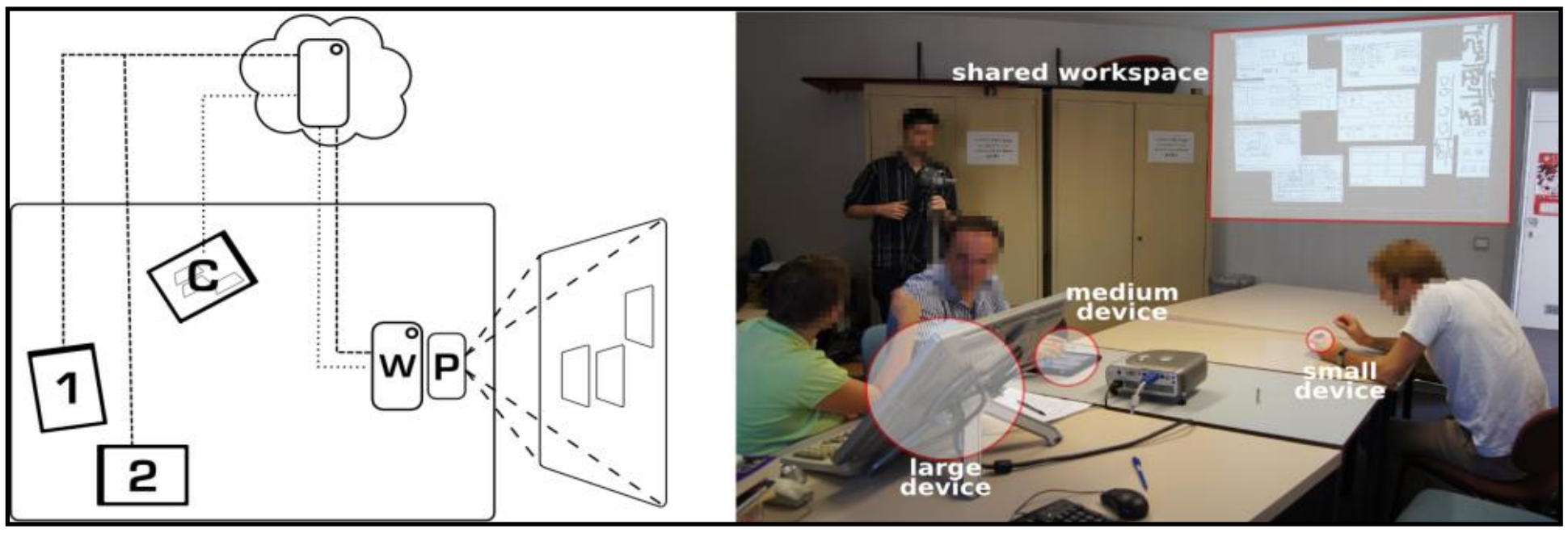

**Figure 7.** Physical setup of GAMBIT.

The collaboration features presented by Gambit are seldom present in other prototyping tools, whether in commercial or academic context. There are some other applications in GUI outside domains that provide similar features, but none of them is applicable in the prototyping domain. Tools like JBart, Axure, Visio, PowerPoint and JustInMind support more simplified mechanisms of collaboration using chat or common repositories, but rarely supporting multiple devices.

It is interesting to notice that many web-based prototyping tools such as Balsamiq, Vectr, Atomic.io, and Proto.io present collaborative features. In addition to functions for editing the prototypes directly in the web browser, they offer services such as a repository to store the prototypes and mechanisms for sharing executable versions of the prototypes with other users. This architecture is well suited for collaborative work since any collaborator can work remotely on the same synchronized repository while maintaining the availability of the prototype for any user who wants to test the prototype.

Some tools like Invision provide mechanisms to manage collaborations among people involved in the process including features for inviting collaborators, supporting discussions, and even assigning tasks. These mechanisms have evolved in the more recent version of Invision released in October 2016 making the team management compatible with the projects tracking tool JIRA. With this new approach, Invision brings the prototyping process closer to the development of the final application itself.

### 3.6. Reuse Mechanism

Reuse is the process of creating software systems from predefined software components. Reusing components previously built is an important feature to promote productivity in software development as they might reduce the workload of designers and users by offering standard UI design. Simple mechanisms to promote reuse might include libraries of widgets, templates and pre-defined behaviors.

Nonetheless, other mechanisms of reuse might be available in specialized tools. For example, sketching tools that support shapes recognition like SILK and SketchiXML offer mechanisms for reusing user-defined drawings that have been previously "trained" by users.

Commercial tools like Appery.io, HotGloo, iRise, Protoshare and UXPin feature the usage of breakpoints and screen version, thus promoting reuse of design for multiple devices. This method consist in creating one version for a screen for each size desired and define breakpoint where the prototype have to switch from one version to another. The advantage of this method is that the prototype can dynamically change completely its layout when resizing the prototype in a preview mode, for instance [24] [25]. This feature is particularly useful when designing a prototype (typically for a website site) that should run on diverse devices (ex. tablet, smartphone or desktop). Some prototyping tools help the user by managing the different versions of a screen, instead of letting the user do it manually. Layouts are therefore completely independent from one breakpoint to another. RWD Wireframe is one of those specialized tools that is dedicated to the management of prototype versions for different screen sizes allowing users to sketch different layout of the prototype for each resolution.

Some tools like ForeUI or MockupScreens allow the reuse of themes. By doing so, it is possible to switch from a wireframe prototype that looks like a sketch to a prototype with the appearance of a real software (e.g. Windows Theme, Mac Theme, etc.) without having to recreate the prototype.

### 3.7. Scenario Management

Scenario-based design is a family of techniques that uses narratives ad scenarios for describing expected outcomes for the system. Narratives are written in very early phases of the development process, and then used to guide both prototyping and the subsequent development of the system [26].

Scenario Management refers to the ability of tools to work with different scenarios and manage them in an integrated way with prototypes and behavior descriptions. It is not an easy feature to implement because it is strongly dependent of the whole development process and their models, so their implementation becomes normally too restrictive. Despite the fact that this feature has appeared first in Freeform in 2003 as a Visual Basic 6 plugin, there has not been much evolution since then. Most of prototyping tools in our survey support scenarios management through simple annotations.

However, we have not found any tool that implements truly scenario management, which might include requirements specifications and tracking decisions along the process.

### 3.8. Preview Mode

Preview Mode is an important feature to allow visualization of an executable version of the prototype. In that mode, we can execute and simulate all interactions specified during the construction of the prototype. Users can test the application as a rough final product. It is important, in this case, to visualize how the prototype will appear in a real environment, perhaps promoting usability testing and collecting adequate feedback from special stakeholders.

MIRAGE, Lapidary and SILK are examples of tools that embed a Preview Mode. DENIM and SketchiXML provide a preview mode with the help of a kind of plugin and/or external tool. All commercial tools provide also some kind of feature to allow execution during development.

An emerging feature called "prototype mirroring" can be understood as a kind of previous mode. Prototype mirroring is implemented by some tools, such as Atomic.io or Origami, that host prototypes on the cloud. This technique allows people to visualize the edition of prototypes in real time using a smartphone (using a proprietary viewer application) and/or a web browser.

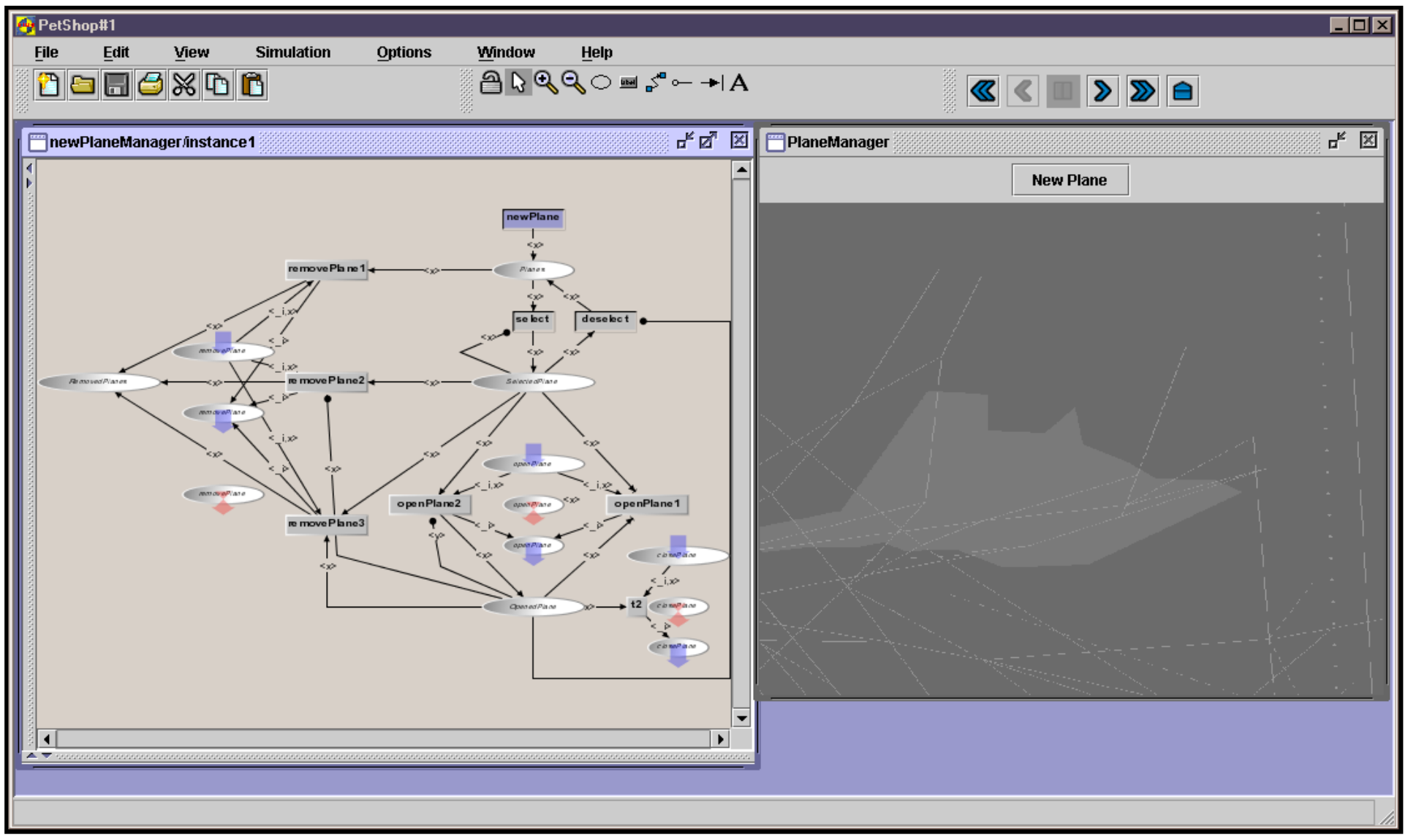

**Figure 8.** Execution of ICO specification in PetShop.

Interactive prototyping, on the other hand, is provided by model-based tools to support co-execution between models and interfaces. Within PetShop [27], for instance, prototyping from specification is performed in an interactive way. At any time during the design process, it is possible to introduce modifications in the models. The ad-vantage of model-based prototyping is that designers can change the model and immediately test the impact on the behavior of the prototype. At run time, the user can both look at the specification and the actual application. Both of them are in two different windows overlapping in Figure 8. The window PlaneManager corresponds to the execution of the window with the Object Petri net underneath.

## 3.9. Annotations

Annotations of prototypes offer the possibility to add informative notes for specific sections of a given artifact. The annotation system is an interesting feature since it may be a way to collect user feedbacks when presenting a prototype to end-users. Users can annotate the prototype to report problems, to indicate preferences about design options, to request clarifications about the design, and to specify parts of the prototype that are not supported by the tool (for example the expected behavior for an animation). Annotations are often meant to be read by other members of the development team for that they should written in a way that it is understandable by others. Naghsh [28] has suggest that annotations can help to create a dialogue and encourage users to participate in the design process.

We have identified three different stages where annotation system is available: Prototype Building, Annotation Mode and Usability Testing. The first and more common stage where the annotation system is available is at the construction of the prototype. At this stage, we have identified two kind of annotations: annotations as a widget and annotations as a property.

Some tools like inPreso (Figure 9) provide widgets dedicated to create annotations. The behavior widgets for annotation is the same of other widgets used to build the prototype (they have properties; they can be resized or moved on the prototype). The most frequent widgets for annotations include callouts, post-its, scratch-outs and arrows. Using widgets, prototypes can be visualized as an annotated document. Annotations as a property is less visible and less pervasive. While widgets or pages of a prototype have their own properties, some tools add a "Note" property where the user can add some text.

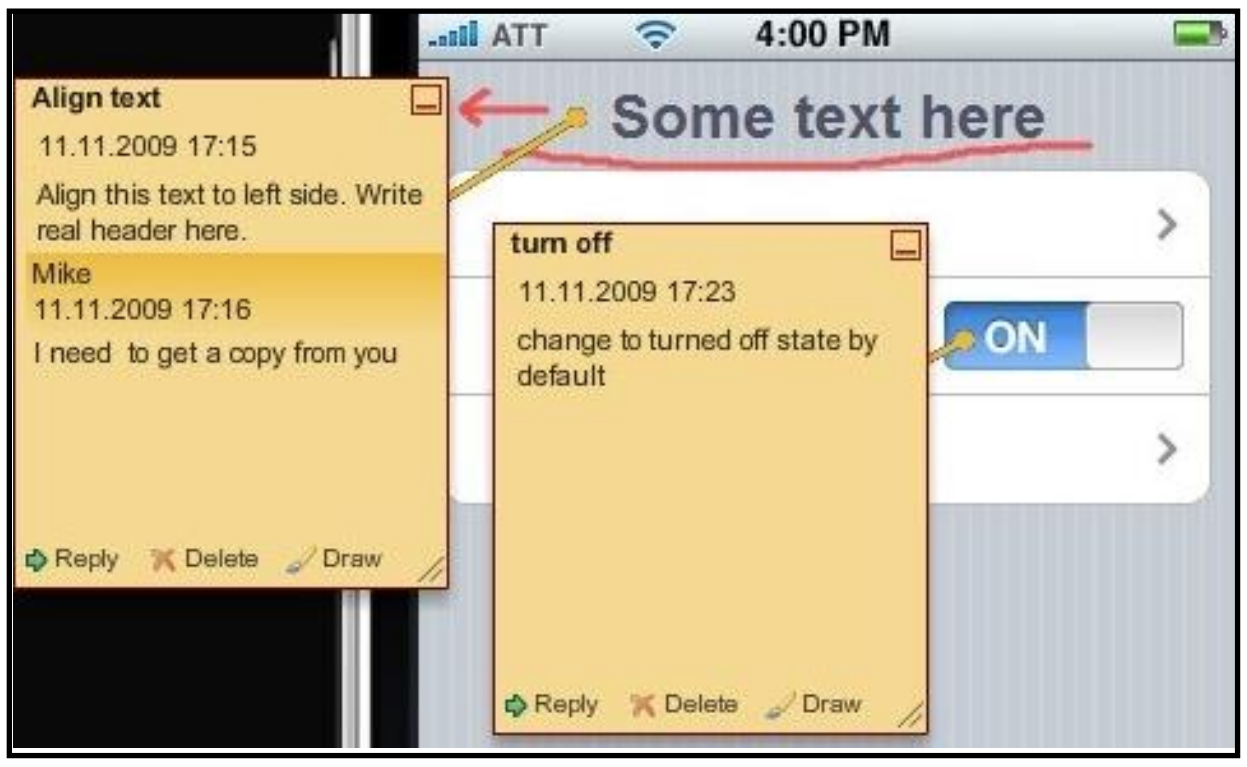

**Figure 9.** Example of a textual annotation using inPreso.

The second stage refers to the annotation mode of the prototype. Indeed, some tools provide a dedicated mode to the annotation system. While it is not possible to modify directly the prototype in this mode, it is possible making annotations or drawing directly on top of the prototype when activating the annotation mode using tools. These tools can be a freehand sketching, a token that can be placed on the prototype with an associated note, or an area that is selected using the mouse.

Annotation mode can also be used during a preview of the prototype. Indeed, once a version of the prototype is finished, it is possible to share it using a link. Any person having the link can test the prototype and make annotations on it. Once the annotation is made, a notification is sent to the person in charge of the prototype.

The last stage refers to the test of the prototype. It is also possible to collect data from users who test the prototype and use it as annotations. Indeed, any information that can be measured while using the prototype (time spent on each screen, the area clicked, etc.) can be saved for a further analysis. This usage is more specific for usability tests, where tools like Solidify provide functionalities that can be useful for that. For instance, it is possible adding instructions or questions to the test of the prototype and creating tasks that have to be accomplished.

SILK and DEMAIS support textual annotations as an input design vocabulary. Some other tools like Alouka, Balsamiq, inPreso, Lumzy and WireFrame Sketcher support annotations through widgets (the simplest method), and others like Axure, Mockup-Screens and JustInMind support this feature as a property. There are also those that have a dedicated annotation mode like Concept.Ly, ForeUI and NinjaMock. However, no tool ensures the annotation system on the three stages at the same time.

## 3.10. Support for Usability Testing

During a typical user testing of a prototype, participants will complete a set of tasks while observers watch, listen and take notes. Any information that can be measured while using the prototype (time spent on each screen, the area clicked, etc.) is worthy of collecting for further analysis. For that, some prototyping tools include functionalities for recording metrics of use.

In addition to annotations, some tools like Solidify and CogTool allow adding instructions to guide users during the use of the prototype. These instructions are presented as questions and/or tasks that are displayed to the participant of the usability test. Users have the possibility to use the prototype to complete tasks, answer the questions or skip them altogether if they are not able to figure out what to do. The tool records the user test and makes the results available through the means of automated annotations of the prototype. These functionalities allow automating the test and making it available as remote surveys.

Using the date collected by tools that support usability testing of prototypes, designers can analyze the click flow, checking statistics for each page as well as demographic filters when displaying the results (Figure 10). These results are useful to support decisions between several designs choices. Some tools like PickFu or IntuitionHQ also provide an interface to plan tests and manage the results.

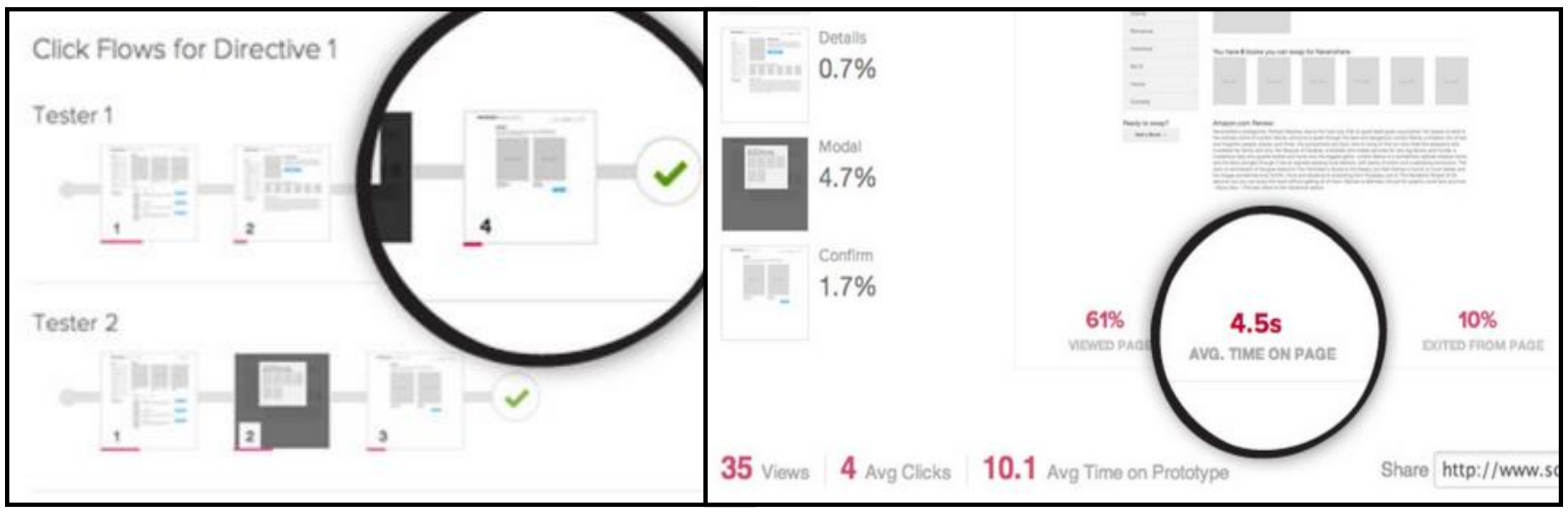

**Figure 10.** Example of data from usability testing collected by Solidify.

While some tools embed mechanisms for usability testing, other tools such as Invision (Figure 11), Marvel, Flinto, Axure, Justinmind and Proto.io provide mechanisms to link the prototype with third-party tools that are specialized in automating the usability test such as UserTesting, Validately or Lookback. For example, UserTesting is a service that provides users feedback on an application, a website or a prototype. They also provide support for running tests, registering recordings (i.e. video, interactions) and analyzing the results. The interest in usability testing is quite recent. Indeed, we can notice for instance that Invision has announced their compatibility with UserTesting on September 2016 or JustInMind announced its partnership with Validately on February 2016.

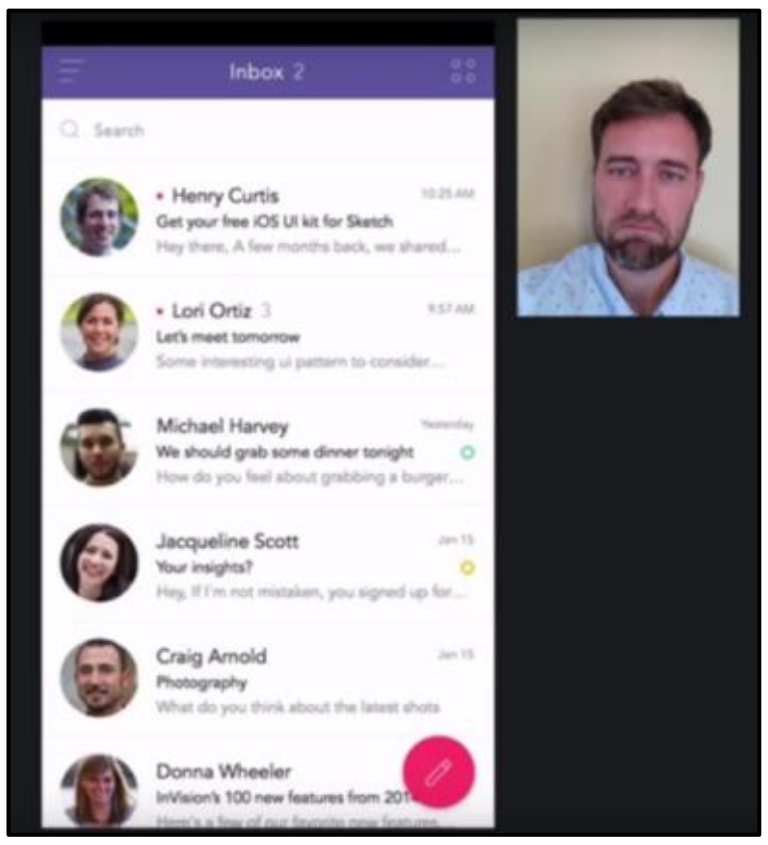

**Figure 11.** Example of a user testing recording with Invision.

## 3.11. Support for Code Generation

Code Generation refers to the capacity of the prototyping tool to produce the code of the final application from a model specification. Code generation can only produce full-fledge applications if the prototyping tools support modeling of both presentation and behavior aspects. The generated code might serve as the basis to develop a final and concrete user interface as well as an exportable output exploitable by other tools. Such is the case of SILK, which generates code for an old OpenLook Operating System, and Freeform, which generates code for Visual Basic 6. SketchiXML and Gambit produces interface specifications and generates code in UsiXML, an open source format based on XML.

Among all commercial tools is our survey, 25 of them can generate web pages based on the prototype. Tools such as AppSketcher, Axure, ForeUI and JustInMind generate web pages that include in the code annotations of the dialogue specification, so that it is possible to reuse these web pages to reengineering the prototype and make it to evolve to the final user interface.

## 3.12. Version Control

Version control is the mechanism that allow development teams to track the evolution of artifacts over time. It allows to answer questions such as how many different/alternative versions exist, what is the current state of the development, and in some cases, the rationale of modifications. Version control is important because prototypes are constantly evolving along the development process to accommodate users' feedback and/or to include new requirements that emerge along the way. Moreover, many prototypes might be produced to explore alternative design options. When alternative options are at stake, it might be necessary to compare two (or more) alternative versions in order to identify the differences.

Alouka (Figure 12), Codiqa, FluidUI, HotGloo and JustInMind support version control. Concept.Ly is able to compare two different screens using a slider. However, it is not possible to compare two versions of one screen, but only two different screens from the same version. SILK supports version control with design history.

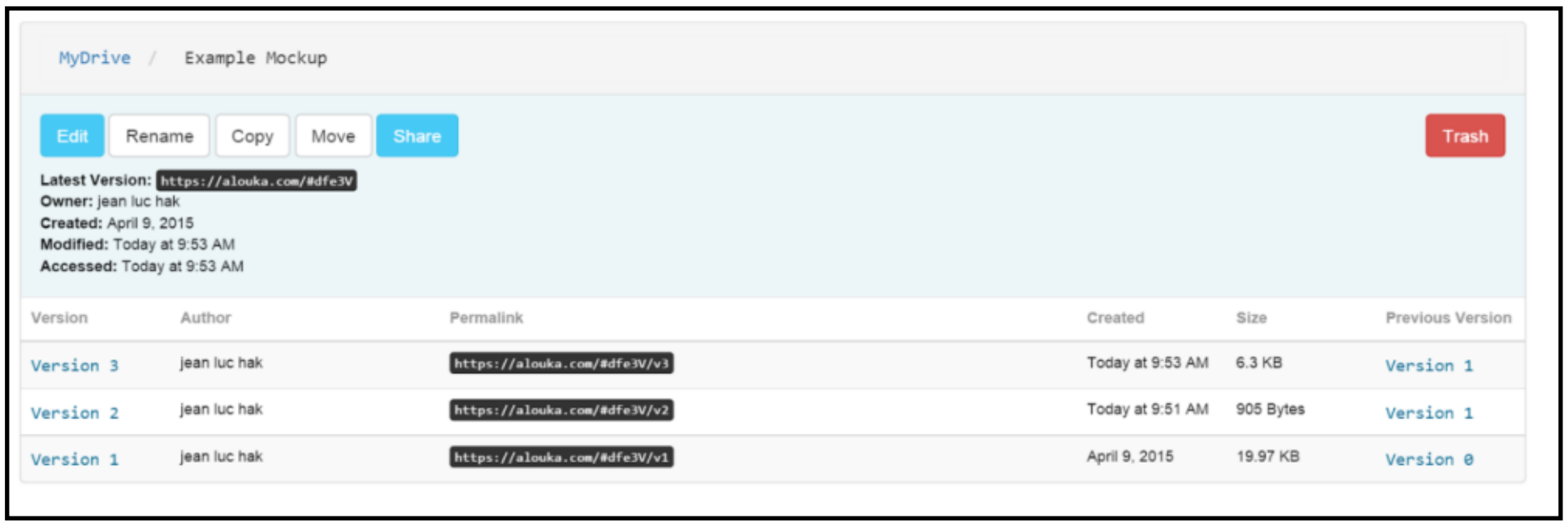

**Figure 12.** Versioning using Alouka.

## 3.13. Support for UCD interactive development

According to the ISO 13407 standard [1], a User-Centered Design (UCD) process features an iterative lifecycle that is meant to guide the development team from phases of requirements engineering, passing by cyclic phases of production and evaluation of design solutions until prototypes evolve into implementations that reach the maturity level required for delivery to the end-users. Since prototyping is one of the core activities in a UCD process, we might expect that prototyping tool should help the development team along all phases.

Seffah and Metzker [29] stressed the need for "computer-assisted usability engineering" tools and frameworks to share best practices between software engineering and user-centered design. UIMS tools might be considered a timid attempt to provide an integrated design solution with emphasis on automation of the GUI building. However, there is an important gap since most of current tools support only "produce design solutions", not giving support for all UCD phases.

GRIP-it is a tool that focuses on the transition of prototypes into the software devel-opment by providing integrated and interoperable tools that help to propagate information about the design among all people involved in the process.

SILK supports the transformation process of the sketches to real widgets and graphical objects, but other steps in the process are not covered. Other sketching platforms such as SketchiXML and Gambit require the integration with third-party UsiXML tools to support several levels of prototyping.

DENIM and DEMAIS do not support different refinement levels, so they do not cover the whole lifecycle (they do not produce finished HTML pages, for example). DENIM just allows navigating among different representations in a web-design prototype, such as site maps, storyboards and mock-ups. Some tools like ScreenArchitect support model description that it is good to provide links between prototypes and models like state machines, leading then to a more integrated environment in UCD development processes.

## 4. Discussion of the findings

In this section, we present a broader analysis of the tools with respect to the milestones. Figure 13 presents a historical view of tools and milestones. We start by classifying tools per year of (first) release. In the case of academic tools, we considered the year of publication. For commercial tools, we sought the year of first appearance in the market. The graph presented at Figure 13 shows the total number of tools released per year and the first occurrence of milestones observed in tools. We have classified tools and milestones in three main periods that roughly cover first attempts for building prototyping tools, for supporting the development process, and the emergence of tools supporting collaborative work.

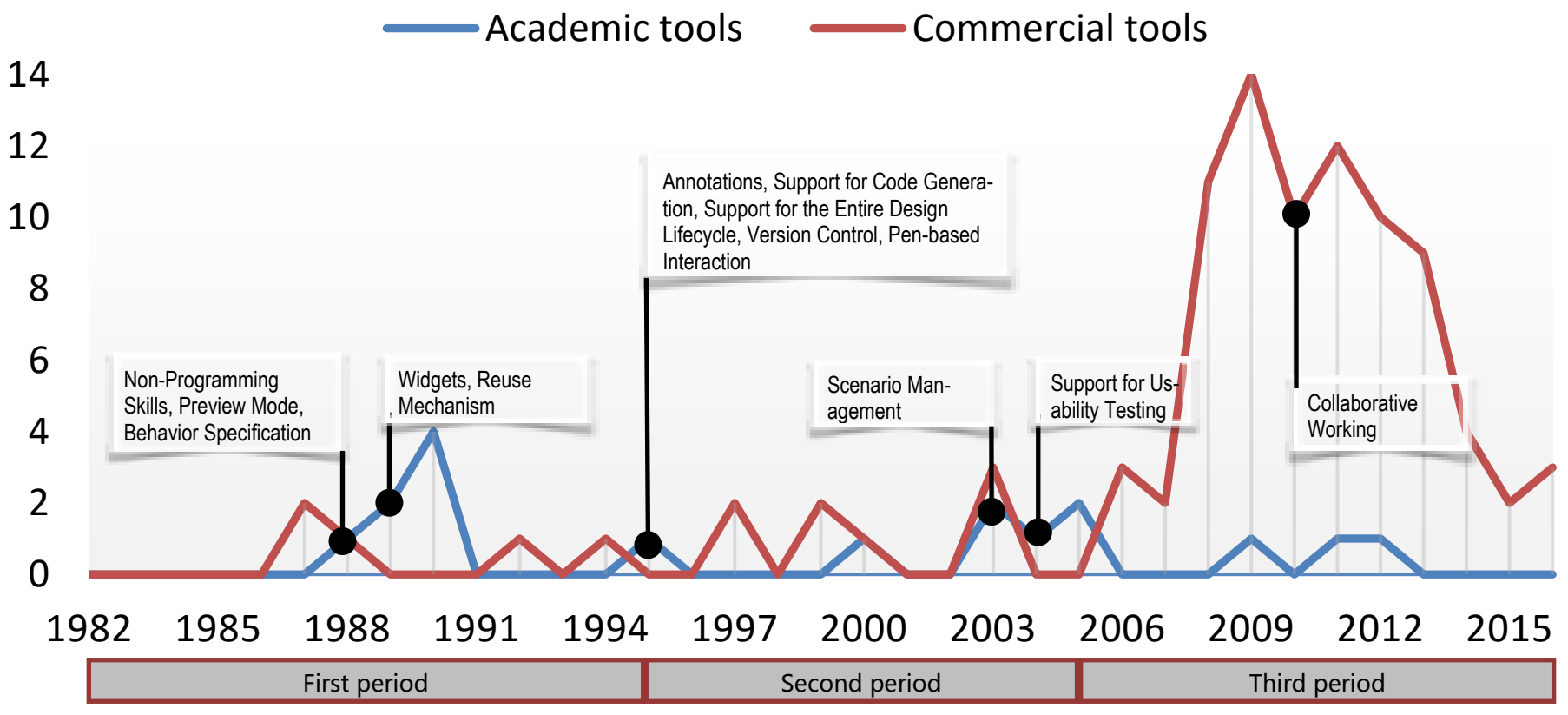

**Figure 13.** Number of both academic and commercial tools per year.

The first period (< 1995) is characterized by the emergence of UIMS tools. Authors claimed that the main advantage of UIMS tools is in the fact that after development and testing, interface prototypes could be attached directly to the application, thus the prototype becomes the industrial interface [4]. UIMS tools focus on high-fidelity proto-types, using mostly design elements from the final interface, and being strongly dependent on the platform. UIMS tools lack the flexibility needed in the early phases of the development process when designers should focus on problems to be solved in terms of business and users' requirements rather than terms of user interface design. In this period, we have also found many reports of using tools such as PowerPoint and Visio to create prototypes. Although PowerPoint and Visio are not intended to build prototypes, they provide functions for drawing presentations and creating transitions, which might have been helpful to build low-fidelity prototypes when no other UIMS tool was available.

The second period (1995-2005) encompasses tools with functionalities to support the development team when managing prototyping activities (ex. annotations, code generation, version control, etc.). There was an increasing interest in the period on alternative ways of prototyping user interfaces as well as in behavior modeling. For example, we observed the emergence of sketching tools such as SILK and DENIM.

The third and last period is characterized by a substantial increase of commercial tools and support for collaborative work. This period goes from 2007 to now.

Along these periods, features like Non-Programming Skills, the use of Widgets and Behavior Specification were the three most implemented by tools (over 70%). This fact can signalize the focus in providing a friendly environment for non-technical people since the first years. McDonald et al. [4] in 1988 had already pointed the need to consider different skills from the various stakeholders involved and to allow they use tools to design their own interfaces without technical skills. The way tools started providing that - and still remain until now - was through Widgets. Widgets have introduced a simple mechanism to encapsulate an idea (and sometimes behaviors) for each component normally used to build GUIs.

Features like Scenario Management, Support for Usability Testing and Support for the Entire Design Lifecycle are supported by a few tools (less 10%). This number suggests a slow progress towards the support of the whole lifecycle of prototyping.

Concerning Pen-Based Interaction, only 9.92% of tools implement this feature. Pen-Based Interaction feature was presented in SILK in 1995, and after some years, well-known tools like Adobe Illustrator and Photoshop implemented it. Nevertheless, it never seems to become a successful feature with commercial prototyping tools. This might be explained by the fact that sketches are hard to maintain (ex. ambiguity of sketches) and hard to make them evolve throughout the development process.

Table 2 summarizes the findings showing a list of all tools retained for analyses in the three periods, ordered by year of launch (the sign of "?" means that was not possible to determine the year of launch), and the set of milestones that each one covers. It also shows the percentage of tools that covers each milestone individually.

**Table 2.** Set of milestones observed per tool.

| Tool | Year | 1 | 2 | 3 | 4 | 5 | 6 | 7 | 8 | 9 | 10 | 11 | 12 | 13 |
|---|---|---|---|---|---|---|---|---|---|---|---|---|---|---|
| iPhoneMockup | ? | | | | | | | | | | | | | |
| iRise | ? | ■ | | ■ | ■ | ■ | | | ■ | | | | ■ | |
| JBart | ? | | | ■ | ■ | ■ | | | | | | ■ | | ■ |
| Mockup Designer | ? | ■ | | ■ | | ■ | | | | | | | | |
| Omnigraffle | ? | ■ | | ■ | ■ | | | | | | | | | |
| ProcessOn | ? | ■ | | ■ | | ■ | ■ | | | | | | ■ | |
| Protostrap | ? | ■ | | ■ | ■ | | ■ | | ■ | | | ■ | | |
| Serena Prot. Composer | ? | ■ | | ■ | ■ | | | | ■ | | | | | |
| SoftAndGUI | ? | ■ | | ■ | ■ | | ■ | | ■ | | | | | |
| UXPin | ? | ■ | | ■ | ■ | ■ | ■ | ■ | ■ | ■ | ■ | | ■ | |
| Adobe XD | ? | ■ | | ■ | ■ | ■ | ■ | | ■ | | | | ■ | |
| Adobe Illustrator | 1987 | ■ | ■ | ■ | | | ■ | | | | | | | |
| Microsoft PowerPoint | 1987 | ■ | | ■ | | | ■ | ■ | ■ | | | | | |
| Adobe Photoshop | 1988 | ■ | ■ | ■ | | | ■ | | | | | | | |
| Mirage | 1988 | ■ | | ■ | | | | | | | | | | |
| Ensemble | 1989 | ■ | | ■ | | | | | | | | | | |
| Lapidary | 1989 | ■ | | ■ | ■ | | | | ■ | | | | | |
| Druid | 1990 | ■ | | ■ | | | | | | | | | | |
| SCENARIOO | 1990 | ■ | | ■ | ■ | | | | ■ | ■ | ■ | ■ | | |
| MoDE | 1990 | ■ | | ■ | ■ | | | | ■ | | | | | |
| SIRIUS | 1990 | | | | ■ | | | | ■ | | | | | |
| Microsoft Visio | 1992 | ■ | | ■ | ■ | ■ | | | | | | | | |
| SmartDraw | 1994 | ■ | | ■ | | | ■ | | | | | | | |
| SILK | 1995 | ■ | ■ | | ■ | | | | ■ | | | | ■ | ■ |
| Adobe Fireworks | 1997 | ■ | | ■ | | | ■ | | | | | ■ | | |
| Micr. Visual Studio | 1997 | | | ■ | ■ | | | | ■ | | | ■ | | ■ |
| Adobe InDesign | 1999 | ■ | ■ | ■ | | | ■ | | | | | | | |
| AutoIt | 1999 | | | ■ | ■ | | | | | | | | | |
| ScreenArchitect | 2000 | | | ■ | ■ | | | ■ | ■ | | | ■ | | ■ |
| DENIM | 2000 | ■ | ■ | | ■ | | | | ■ | | | ■ | | |
| Axure | 2003 | ■ | | ■ | ■ | ■ | ■ | | ■ | | | ■ | | |
| Inkscape | 2003 | ■ | ■ | | | | | | | | | | | |
| KeyNote | 2003 | ■ | | ■ | | | ■ | ■ | ■ | | | | | |
| DEMAIS | 2003 | ■ | ■ | | | | | | | | | | | |
| Freeform | 2003 | | ■ | | | | | ■ | | | | | | |
| CogTool | 2004 | ■ | | ■ | ■ | | ■ | ■ | ■ | ■ | ■ | | | |
| SketchiXML | 2005 | ■ | ■ | | | | ■ | | ■ | | | ■ | | |
| Monet | 2005 | ■ | ■ | | ■ | | | | | | | | | |
| GUI Design Studio | 2006 | ■ | | ■ | ■ | ■ | ■ | ■ | ■ | | | | ■ | |
| JotForm | 2006 | ■ | | ■ | ■ | | | | ■ | | | ■ | | |
| MockupScreens | 2006 | ■ | | ■ | ■ | | ■ | ■ | ■ | | | ■ | | |
| JustInMind | 2007 | ■ | | ■ | ■ | ■ | ■ | ■ | ■ | ■ | ■ | | ■ | |
| Micr. Expression Blend | 2007 | | | ■ | ■ | | ■ | | ■ | | | | | ■ |
| Balsamiq | 2008 | ■ | | ■ | ■ | | | | ■ | | | | | |
| ConceptDraw | 2008 | ■ | | ■ | | | | | | | | | | |

| | |
|---|---|
| DesignerVista | 2008 |
| inPreso Screens | 2008 |
| Matisse (Swing GUI B) | 2008 |
| MockingBird | 2008 |
| Pencil project | 2008 |
| Pidoco | 2008 |
| ProtoShare | 2008 |
| PickFu | 2008 |
| WireFrameSketcher | 2008 |
| ActiveStory Enhanced | 2009 |
| Cacoo | 2009 |
| Crank Storyboard Des. | 2009 |
| Creately | 2009 |
| DevRocket | 2009 |
| FlairBuilder | 2009 |
| ForeUI | 2009 |
| Gliffy | 2009 |
| GUI Machine | 2009 |
| LovelyCharts | 2009 |
| Microsoft Sketchflow | 2009 |
| Napkee | 2009 |
| IntuitionHQ | 2009 |
| iPlotz | 2009 |
| Simulify | 2009 |
| Adobe Flash Catalyst | 2010 |
| Appery.io | 2010 |
| BluePrint | 2010 |
| FrameBox | 2010 |
| HotGloo | 2010 |
| LucidChart | 2010 |
| MockaBilly | 2010 |
| Mockflow | 2010 |
| Naview | 2010 |
| Sketch | 2010 |
| 10Screens | 2011 |
| Antetype | 2011 |
| AppCooker | 2011 |
| Draw.io | 2011 |
| FieldTest | 2011 |
| InsitUI | 2011 |
| Lumzy | 2011 |
| MockupBuilder | 2011 |
| Mockups.me | 2011 |
| Mockup Tiger | 2011 |

| | | | | | | | | | | | | | | |
|---|---|---|---|---|---|---|---|---|---|---|---|---|---|---|
| PowerMockup | 2011 | ■ | | ■ | | | | | | | | | | |
| Proto.io | 2011 | ■ | | ■ | ■ | ■ | ■ | | ■ | ■ | ■ | ■ | ■ | ■ |
| GRIP-it | 2011 | ■ | | | | | | | | | | | | ■ |
| AppMockupTools | 2012 | ■ | | ■ | | | ■ | | | | | | | |
| AppSketcher | 2012 | | | | ■ | | | | | | | | | |
| Codiqa | 2012 | ■ | | ■ | ■ | ■ | | | ■ | | | | ■ | |
| FluidUI | 2012 | ■ | | ■ | ■ | | | | ■ | | | ■ | ■ | |
| Indigo Studio | 2012 | ■ | | ■ | ■ | | ■ | ■ | ■ | | | | | |
| Moqups | 2012 | ■ | | ■ | ■ | | ■ | | ■ | | | | ■ | |
| Prototyping On Paper | 2012 | ■ | | | ■ | | ■ | | ■ | | | | | |
| SceneBuilder | 2012 | | | ■ | ■ | | | | ■ | | | ■ | | ■ |
| Solidify | 2012 | ■ | | | ■ | | ■ | ■ | ■ | ■ | ■ | | | |
| FrameJS | 2012 | ■ | | ■ | ■ | ■ | ■ | | ■ | | | ■ | | |
| Gambit | 2012 | ■ | ■ | | | ■ | | | ■ | | | | | |
| Alouka | 2013 | ■ | | ■ | | | ■ | | ■ | | | | | |
| Concept.ly | 2013 | ■ | | | ■ | ■ | ■ | | ■ | | | | ■ | |
| Flinto | 2013 | ■ | | | ■ | ■ | ■ | | ■ | | | | ■ | |
| InVision | 2013 | ■ | | | ■ | ■ | ■ | | ■ | ■ | ■ | | ■ | ■ |
| Marvel | 2013 | ■ | | | ■ | | ■ | | ■ | | | ■ | | |
| NinjaMock | 2013 | ■ | | ■ | ■ | | ■ | | ■ | | | ■ | | |
| Notism | 2013 | ■ | | | ■ | ■ | ■ | | ■ | | | | ■ | |
| RWD Wireframes | 2013 | ■ | | ■ | | | ■ | | | | | ■ | | |
| Webflow | 2013 | ■ | | ■ | ■ | ■ | ■ | | ■ | | | ■ | ■ | |
| AppGyver Prototyper | 2014 | ■ | | ■ | ■ | | ■ | | ■ | | | | | |
| Avocado | 2014 | ■ | | | ■ | | | | ■ | | | | | |
| Mockup Plus | 2014 | ■ | | ■ | ■ | | ■ | | ■ | | | | | |
| SnapUp | 2014 | ■ | | | ■ | ■ | ■ | | | | | ■ | | |
| Atomic | 2015 | ■ | | ■ | ■ | ■ | ■ | | ■ | | | | ■ | |
| Easee | 2015 | | | ■ | | | ■ | | ■ | | | ■ | | |
| Principle | 2016 | ■ | | ■ | ■ | | | | ■ | | | | | |
| Vectr | 2016 | ■ | | ■ | | ■ | ■ | | | | | | | |
| Origami | 2016 | ■ | | ■ | ■ | | ■ | | ■ | | | ■ | | |
| | **Total: 121** | 105 | 12 | 93 | 85 | 34 | 71 | 12 | 81 | 51 | 10 | 33 | 29 | 13 |
| | **Percentage:** | **86** | **9,9** | **76** | **70** | **28** | **58** | **9,9** | **66** | **42** | **8,2** | **27** | **23** | **10** |

Figure 14 presents a graph with the percentage of milestones covered by tools. We can notice that the five more covered milestones (Non-Programming Skills, the use of Widgets, Behavior Specification, Preview Mode and Reuse Mechanism) – all of them covered by more than half of tools – are also the oldest features presented by prototyping tools (since 1988). However, the availability of features like Behavior Specification, Preview Mode and Reuse Mechanism evolved along the time. Behavior Specification has benefited from more human-centered approaches such as Scenario-based specifications, while Preview Mode has incorporated co-execution between models and prototypes like in PetShop [27] and ScreenArchitect. Since 2001, Reuse Mechanisms started to include technics like Plastic Interfaces [24] and Responsive Design [25].

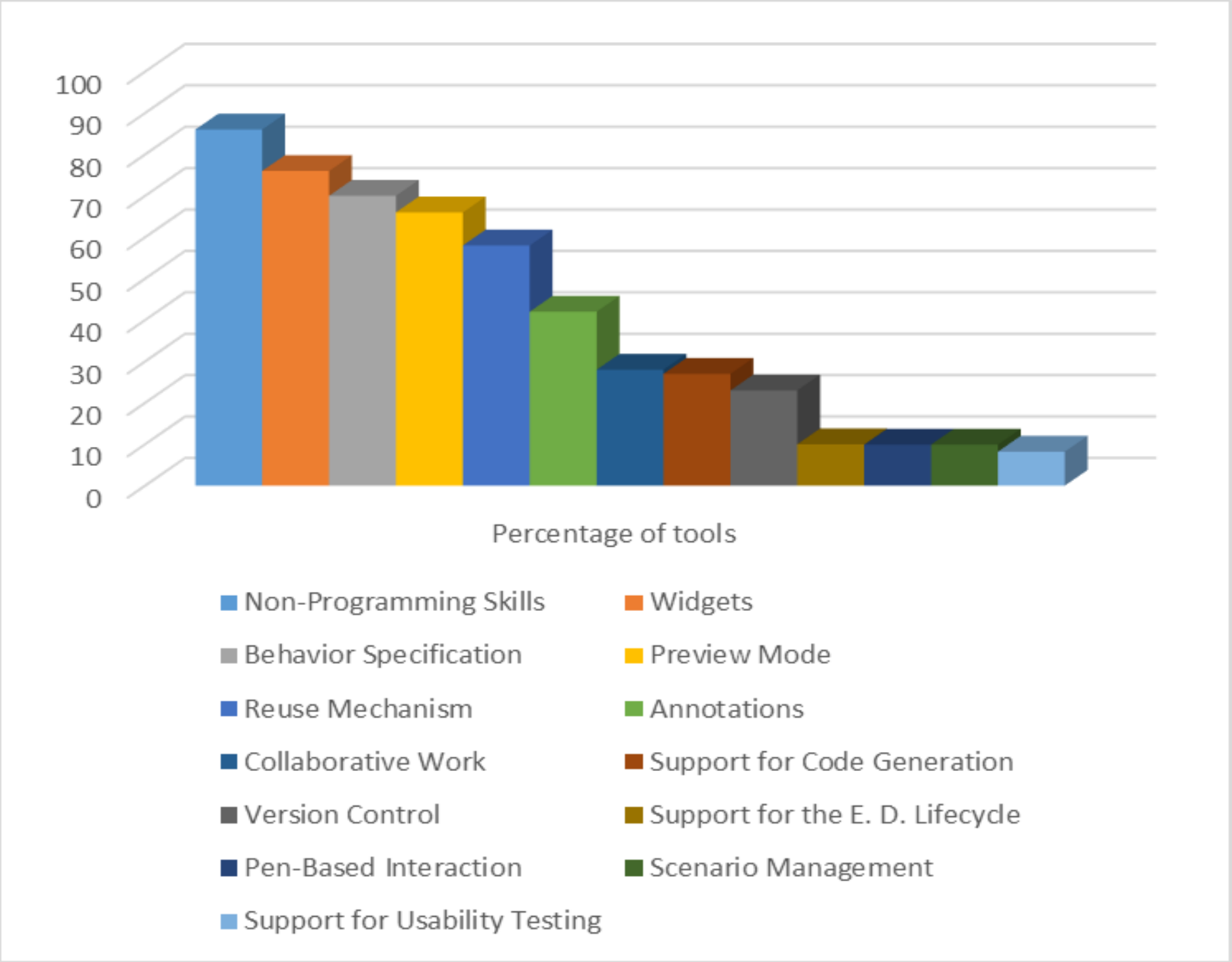

**Figure 14.** Percentage of milestones cover by the analyzed tools.

## 5. Conclusion

This paper presents a survey of academic and commercial tools. The analysis of these tools allowed us to identify some milestone that help to characterize the progress and the evolution of prototyping tools over time.

The analysis of commercial tools is important because their adoption of features has an impact of the practice in the industry. Quite often, academic tools are pioneer in proposing new features that only appear later on commercial tools. In our study however, we did not collect information for analyzing the occurrence of a technological transfer. Many of the innovative features come first from academic tools. However, if the temporal appearance of tools might suggest a possible transfer from academic work to the industry, the present work cannot clearly determine whether (or not) that transfer really occurred. However, we can say that some features like Pen-based Interaction, which were already present in early academic works (SILK, 1995) twenty years ago, did not make so far a breakthrough to commercial tools.

Another aspect we can highlight in this study is the number of commercial tools released after 2008. These tools have incorporated the most aspects we report in this paper, providing, in different levels, implementations of these concepts, and many times, being strongly repetitive in their qualities. Nevertheless, it shows a continued interest both from academic and industrial communities in this theme, suggesting an open space of research in several points. The number of commercial tools also suggests the existence of a market and an increase interest in this type of tool.

Future directions for research point to the development of tools for prototyping as support activity for the development lifecycle. Regarding this gap, we have identified little support of tools for annotation activities in a requirements process. Tools that treat annotations as a property and not as a single remark support a better specification process for gathering requirements. Even though, the way they capture the information coming from those annotations is not profitable to be used for supporting business rules, specification of needs or more formal functional descriptions.

Another important gap identified is related to integrated support for development models. Task and system models, for example, are only considered by few tools. Developing incremental prototypes requires an integrated environment supporting specification of scenarios, models and constraints. Potential tools should consider providing such environment where prototypes could be fully specified, modeled, run and tested.

The analysis presented in this work provides us insights about the drawbacks of existing prototyping tools. In particular, this analysis pinpointed the lack of support for a rationale design and for tracking the decisions made along the development process. Currently, we are working on a tool support called PANDA (Prototyping using Annotation aNd Decision Analysis) [31].

## References

[1] ISO 13407 (ISO 9241-11). (1998). Ergonomic requirements for office work with visual display terminals (VDTs) Part 11: Guidance on Usability. Revised by ISO 9241-171 (2008).

[2] Snyder, C. (2003). Paper prototyping: The fast and easy way to design and refine user interfaces. Morgan Kaufmann.

[3] Schvaneveldt, R, W., McDonald, J. E., & Cooke, N. (1985). Towards a Modular User Interface. (CRL Technical Report No. MCCS-85-10). Computing Research Laboratory, New Mexico State University, Las Cruces, New Mexico.

[4] McDonald, J. E., Vandenberg, P. D., & Smartt, M. J. (1988). The mirage rapid interface prototyping system. In Proceedings of the 1st annual ACM SIGGRAPH symposium on User Interface Software (pp. 77-84). ACM.

[5] Fenn, J., & Raskino, M. (2008). Mastering the hype cycle: how to choose the right innovation at the right time. Harvard Business Press.

[6] Landay, J. A., & Myers, B. A. (1995). Interactive sketching for the early stages of user interface design. In Proceedings of the SIGCHI CHI (pp. 43-50). ACM Press/Addison-Wesley.

[7] Lin, J., Newman, M. W., Hong, J. I., & Landay, J. A. (2000). DENIM: finding a tighter fit between tools and practice for Web site design. In Proc. SIGCHI CHI (pp. 510-517). ACM.

[8] Bailey, B. P., & Konstan, J. A. (2003). Are informal tools better?: comparing DEMAIS, pencil and paper, and authorware for early multimedia design. In Proceedings of the SIGCHI conference on human factors in computing systems (pp. 313-320). ACM.

[9] John, B. E., Prevas, K., Salvucci, D. D., & Koedinger, K. (2004). Predictive human performance modeling made easy. In Proc. of the SIGCHI CHI s(pp. 455-462). ACM.

[10] Sangiorgi, U. B., Beuvens, F., & Vanderdonckt, J. (2012). User interface design by collaborative sketching. In Proceedings of the DIS (pp. 378-387). ACM.

[11] Van den Bergh, J., Sahni, D., Haesen, M., Luyten, K., & Coninx, K. (2011). GRIP: get better results from interactive prototypes. In Proceedings of the 3rd ACM SIGCHI symposium on Engineering interactive computing systems (pp. 143-148). ACM.

[12] Powers, M. K. (1989). Ensemble: a graphical user interface development system for the design

and use of interactive toolkits. In Proc. of ACM SIGGRAPH 1989 (pp. 168-178). ACM.

[13] Myers, B. A., Zanden, B. V., & Dannenberg, R. B. (1989). Creating graphical interactive application objects by demonstration. In Proceedings of the 2nd annual ACM SIGGRAPH symposium on User interface software and technology (pp. 95-104). ACM.

[14] Singh, G., Kok, C. H., & Ngan, T. Y. (1990). Druid: a system for demonstrational rapid user interface development. In Proc. of ACM SIGGRAPH 1990 (pp. 167-177). ACM.

[15] Li, Y., & Landay, J. A. (2005). Informal prototyping of continuous graphical interactions by demonstration. In Proceedings of the 18th annual ACM symposium on User interface software and technology (pp. 221-230). ACM.

[16] Windsor, P. (1990). An object-oriented framework for prototyping user interfaces. In Proceedings of the IFIP TC13 Third International Conference on Human-Computer Interaction (pp. 309-314). North-Holland Publishing Co.

[17] Yen-Ping Shan. (1990). An object-oriented UIMS for rapid prototyping. In Proceedings of the IFIP TC13 Third International Conference on Human-Computer Interaction (pp. 633-638). North-Holland Publishing Co.

[18] Roudaud, B., Lavigne, V., Lagneau, O., & Minor, E. (1990). SCENARIOO: a new generation UIMS. In Proceedings of the IFIP TC13 Third International Conference on Human-Computer Interaction (pp. 607-612). North-Holland Publishing Co.

[19] Plimmer, B., & Apperley, M. (2003). Software to sketch interface designs. In Ninth International Conference on Human-Computer Interaction (pp. 73-80).

[20] Coyette, A., & Vanderdonckt, J. (2005). A sketching tool for designing anyuser, anyplatform, anywhere user interfaces. In IFIP Conference on Human-Computer Interaction (pp. 550-564). Springer Berlin Heidelberg.

[21] Hosseini-Khayat, A., Ghanam, Y., Park, S., & Maurer, F. (2009). ActiveStory Enhanced: Low-Fidelity Prototyping and Wizard of Oz Usability Testing Tool. In Int. Conf. on Agile Processes and Extreme Programming in Software Engineering (pp. 257-258). Springer Berlin Heidelberg.

[22] Kasik, D. J. (1982). A user interface management system. In ACM SIGGRAPH Computer Graphics (Vol. 16, No. 3, pp. 99-106). ACM.

[23] Olsen, D. (1992). User interface management systems: models and algorithms. Morgan Kaufmann Publishers Inc.

[24] Calvary, G., Coutaz, J., & Thevenin, D. (2001). Supporting context changes for plastic user interfaces: a process and a mechanism. In People and Computers XV—Interaction without Frontiers (pp. 349-363). Springer London.

[25] Marcotte, E. (2014). Responsive web design. Second Edition, 153 p. A Book Apart, LLC.

[26] Rosson, M. B., & Carroll, J. M. (2002). Usability engineering: scenario-based development of human-computer interaction. Morgan Kaufmann.

[27] Navarre, D., Palanque, P., & Bastide, R. (2002). Model-Based Interactive Prototyping of Highly Interactive Applications. In Computer-Aided Design of User Interfaces III (pp. 205-216). Springer Netherlands.

[28] Naghsh, A. M., Dearden, A., & Özcan, M. B. (2005). Investigating annotation in electronic paper-prototypes. In Int. Workshop on Design, Specification, and Verification of Interactive Systems (pp. 90-101). Springer Berlin Heidelberg.

[29] Seffah, A., & Metzker, E. (2004). The obstacles and myths of usability and software engineering. Communications of the ACM, 47(12), 71-76.

[30] Lieberman, H., Paterno, F., and Wulf, V. (eds.) End-User Development. Kluwer/Springer, 2005.

[31] Hak, J.L., Winckler, M., Navarre, D. PANDA: prototyping using annotation and decision analysis. In Proceedings of ACM SIGCHI EICS 2016, Brussels, Belgium, June 21-24, 2016. ACM 2016, ISBN 978-1-4503-4322-0.